\documentclass[superscriptaddress,twocolumn,showpacs,amsmath,amssymb]{revtex4}
\usepackage{graphicx}
\usepackage{color}
\renewcommand{\vec}[1]{\mbox{\boldmath $#1$}}

\begin{document}

\title{
Three-body model calculation of the 2$^+$ state in $^{26}$O}

\author{K. Hagino}
\affiliation{ 
Department of Physics, Tohoku University, Sendai 980-8578,  Japan} 
\affiliation{Research Center for Electron Photon Science, Tohoku University, 1-2-1 Mikamine, Sendai 982-0826, Japan}
%E-mail:hagino@nucl.phygs.tohoku.ac.jp

\author{H. Sagawa}
\affiliation{
RIKEN Nishina Center, Wako 351-0198, Japan}
\affiliation{
Center for Mathematics and Physics,  University of Aizu, 
Aizu-Wakamatsu, Fukushima 965-8560,  Japan}
%E-mail: sagawa@u-aizu.ac.jp


\begin{abstract}
We discuss the energy of the excited 2$^+$ state in the unbound 
nucleus $^{26}$O 
using the three-body model of $^{24}$O+$n$+$n$. This model fully takes 
into account the continuum effects as well as the dineutron correlation 
of the valence neutrons. 
The present calculation yields the energy of 
1.354 MeV, which is 
slightly smaller than the unperturbed energy, 1.54 MeV. 
The energy shifts for the ground and 
the 2$^+$ states with respect to the unperturbed energy suggest that 
0$^+$ and the 2$^+$ states in $^{26}$O show 
a typical spectrum 
well described by a short-range pairing
interaction 
acting on two valence nucleons in the same orbit. 
\end{abstract}

\pacs{21.10.-k,21.60.Cs,23.90.+w,27.30.+t}

\maketitle

One of the most important problems in the physics of unstable 
nuclei is to identify the location of the neutron drip line. 
So far, the drip line has been known experimentally up to oxygen 
isotopes, for which the last bound nucleus is $^{24}$O \cite{S99}. 
An interesting phenomenon has been found experimentally 
in the oxygen isotopes. That is, 
the oxygen ($Z$=8) isotopes are 
abruptly end at $N=16$, while the fluorine ($Z$=9) 
isotopes extend significantly more, at least up 
to $^{31}$F ($N=22$) \cite{S99}. 
Recently, this phenomenon has been successfully explained in terms of 
a three-body interaction \cite{OSHSA10}. 

In order to shed more light on the anomalous behavior of the neutron 
drip line in the oxygen region, it is useful to investigate the properties 
of unbound oxygen nuclei beyond the drip line. 
The neutron decay energy spectrum for the unbound nucleus $^{25}$O 
was studied for the first time by Hoffman {\it et al.}, with which 
the decay energy and the width were determined to be 770$^{+20}_{-10}$ keV 
and 172(30) keV, respectively \cite{H08}. 
These values have subsequently 
been confirmed also by 
Caesar {\it et al.} \cite{CSA13}, 
although the decay width 
was measured to be somewhat smaller, 
that is, 20$^{+60}_{-20}$ keV \cite{CSA13}. 
For the $^{26}$O nucleus, 
its two neutron emission decay from the ground state (that is, 
$^{26}$O $\to$ $^{24}$O + $n$ + $n$) was 
identified for the first time by Lunderberg {\it et al.} \cite{LDK12}, 
which was followed by a measurement 
of Caesar {\it et al.} \cite{CSA13}. 
The decay energy determined in Ref. \cite{LDK12} was 150$^{+50}_{-150}$ keV. 
Later on, Kohley {\it et al.} extracted the decay half-life of $^{26}$O 
to be 4.5$^{+1.1}_{-1.5}\pm 3$ ps \cite{KLD13}. 
These experimental data on the ground state decay of 
$^{26}$O have been theoretically analyzed 
by Grigorenko {\it et al}. \cite{GMZ13} 
as well as by the present authors \cite{HS14}. 
In particular, Grigorenko {\it et al}. have concluded 
based on their three-body model calculation 
that the decay half life of 4.5$^{+1.1}_{-1.5}\pm 3$ ps corresponds to the 
decay energy which is smaller than about 1 keV \cite{GMZ13}. 

In this paper, we discuss the excited 2$^+$ state of $^{26}$O using 
the same three-body model as the one employed in Ref. \cite{HS14}.  
A signal of 
the second peak in the decay spectrum, 
which is likely due to the excited 2$^+$ state based 
on shell model calculations\cite{BR06,VZ06}, 
was not strong in the experimental 
data of Lunderberg {\it et al.} \cite{LDK12} and 
Caesar {\it et al.} \cite{CSA13}, mainly because the statistics were not 
sufficient. 
Nevertheless, 
it is worthwhile investigating the 2$^+$ state in $^{26}$O 
in front of the forthcoming experiment with the SAMURAI facility 
at RIKEN \cite{Kondo}.

A striking fact is that most of theoretical calculations 
performed so far 
 have yielded a larger value of 
the energy of the 2$^+$ state, $E_{2^+}$, 
compared with the unperturbed energy, 1.54 MeV, that is 
twice the energy of the single-particle  d$_{3/2}$ resonance state in $^{25}$O. 
An ab-initio calculation with chiral $NN$ and 3$N$ interactions predicts 
$E_{2^+}$ to be 1.6 MeV above the ground state \cite{CSA13} (see also 
Ref. \cite{HBCLR13}). 
Shell model calculations with the USDA and USDB interactions \cite{BR06} 
yield $E_{2^+}$ to be 1.9 and 2.1 MeV, respectively \cite{CSA13}. 
These theoretical 
results may imply that the continuum effects, which play an 
essential role in unbound nuclei, are not taken into account sufficiently 
well in these calculations. The continuum effects are supposed to be 
included in a consistent manner in the continuum shell model calculation of 
Ref. \cite{VZ06}, but the calculation still overestimates 
the 2$^+$ energy 
to be 1.8 MeV, probably because the ground state energy of $^{25}$O 
is also overestimated by about 230 keV. 
Given this puzzling situation, it is intriguing 
to investigate the 
2$^+$ state in $^{26}$O using the three-body model\cite{HS14}, 
which takes fully into account the continuum effects as well as the 
dineutron correlation between 
the valence neutrons \cite{CIMV84,HS05,MMS05,PSS07,HTS13}. 

In order to describe the decay process of $^{26}$O nucleus, we assume 
the three-body structure of $^{24}$O+$n$+$n$. 
We employ the same Hamiltonian as in Ref. \cite{HS14}, that is, 
\begin{equation}
H=\hat{h}_{nC}(1)+\hat{h}_{nC}(2)+v(1,2), 
\label{3bH}
\end{equation}
in which $\hat{h}_{nC}$ is the single-particle Hamiltonian 
for a valence neutron interacting with the core, and 
$v(1,2)$ is the pairing interaction between the valence neutrons. 
For simplicity, we have neglected the two-body part of the recoil 
kinetic energy of the core nucleus, while the one-body part is included 
in the single-particle 
Hamiltonian $\hat{h}_{nC}$ through the reduced mass. 
The single-particle potential in $\hat{h}_{nC}$ has a Woods-Saxon form, whose 
parameters are chosen in order to 
reproduce the energy of the $d_{3/2}$ resonance state of $^{25}$O 
at 770 keV. For the two-body pairing interaction, $v(1,2)$, we employ 
a density-dependent contact interaction \cite{BE91,HS05,EBH97}, 
\begin{equation}
v(\vec{r}_1,\vec{r}_2)=\delta(\vec{r}_1-\vec{r}_2)
\left(v_0+\frac{v_\rho}{1+\exp[(r_1-R_\rho)/a_\rho]}\right), 
\label{vnn}
\end{equation}
in which 
the first term describes the interaction in the vacuum whereas the 
second term simulates the medium effect through the density dependence of the interaction. 
See Ref. \cite{HS14} for the values of the parameters for the single-particle 
potential and for the pairing interaction. 

With the three-body model, we compute the decay energy spectrum 
for a given angular momentum $I$, 
\begin{equation}
\frac{dP_I}{dE}=\sum_k|\langle\Psi_k^{(I)}|\Phi^{(I)}_{\rm ref}\rangle|^2
\,\delta(E-E_k), 
\label{decayspectrum}
\end{equation}
where $\Psi_k^{(I)}$ is a 
solution of the three-body model Hamiltonian with 
the angular momentum $I$ and the energy $E_k$, 
and $\Phi^{(I)}_{\rm ref}$ is the wave function 
for a reference state with the same angular momentum. 
The reference state can be somewhat arbitrary. In Ref. \cite{HS14}, we have 
used the ground state of the $^{27}$F nucleus obtained by solving a 
three-body model Hamiltonian for $^{25}$F+$n$+$n$. 
In this paper, for simplicity, 
we instead use the uncorrelated state 
of $^{27}$F with the 
neutron $|[1d_{3/2}\otimes1d_{3/2}]^{(IM)}\rangle$ configuration, which is 
dominant in the ground state of $^{27}$F. 
We have checked that both the reference states lead to 
similar decay energy spectra for the ground state decay. 

With a contact interaction, the continuum effects on the decay energy 
spectrum can be easily taken into 
account in terms of the Green's function. In order to demonstrate this, 
first notice that Eq. (\ref{decayspectrum}) can be expressed as 
\begin{eqnarray}
\frac{dP_I}{dE}&=&
-\frac{1}{\pi}\Im 
\sum_k
\langle\Phi^{(I)}_{\rm ref}|\Psi_k^{(I)}\rangle\,\frac{1}{E_k-E-i\eta}\,
\langle\Psi_k^{(I)}|\Phi^{(I)}_{\rm ref}\rangle, \nonumber \\
\\
&=&
-\frac{1}{\pi}\Im 
\langle\Phi^{(I)}_{\rm ref}|G^{(I)}(E)|\Phi^{(I)}_{\rm ref}\rangle, 
\end{eqnarray}
where $\Im$ denotes the imaginary part and $\eta$ is an infinitesimal number. 
In this equation, the correlated Greens's function, $G^{(I)}(E)$, is given 
by 
\begin{equation}
G^{(I)}(E)=
\sum_k
|\Psi_k^{(I)}\rangle\,\frac{1}{E_k-E-i\eta}\,
\langle\Psi_k^{(I)}|. 
\end{equation}
Notice that the correlated Green's function can be constructed 
as \cite{EB92,HS14}, 
\begin{equation}
G^{(I)}(E) = G^{(I)}_0(E)-G^{(I)}_0(E)v(1+G^{(I)}_0(E)v)^{-1}G^{(I)}_0(E), 
\end{equation}
where $G_0^{(I)}(E)$ is the unperturbed Green's function 
given by 
\begin{equation}
G^{(I)}_0(E)=\sum_{\rm 1,2}\frac{|(j_1j_2)^{(IM)}\rangle\langle(j_1j_2)^{(IM)}|}
{e_1+e_2-E-i\eta}. 
\label{green0}
\end{equation}
Here, 
the sum includes all independent 
two-particle states coupled to the total angular momentum of $I$. 
See Ref. \cite{EB92} for a practical method to perform a summation over 
a continuum spectrum for the single-particle states. 

\begin{figure} 
\includegraphics[scale=0.5,clip]{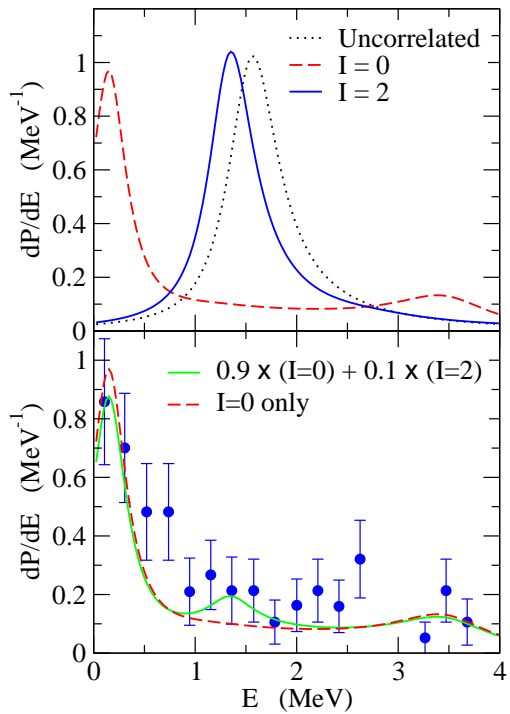}
\caption{(Color online) 
(the upper panel)
The decay energy spectrum for the two-neutron emission decay of $^{26}$O. 
The dashed and the solid lines are for the 0$^+$ and 2$^+$ states, 
respectively. The dotted line shows the uncorrelated spectrum obtained 
by ignoring the interaction between the valence neutrons. 
(the lower panel) 
The decay energy spectrum obtained by superposing the $I=0$ and $I=2$ 
components as is indicated in the figure. The dashed line is the same 
as the one in the upper figure, that is, the decay energy spectrum for the 
pure $I=0$ configuration. 
The experimental data, normalized to the unit area, are 
taken from Ref. \cite{LDK12}. 
}
\end{figure}

The upper panel of Fig. 1 shows the decay energy spectrum 
of $^{26}$O 
for the $I=0$ (the dashed line) and $I$=2 (the solid line). 
For a presentation purpose, we set $\eta$ in Eq. (\ref{green0}) to be a 
finite value, that is, $\eta$=0.21 MeV \cite{HS14}. 
For comparison, 
we also show the spectrum for the uncorrelated case by the dotted line, 
which gives the same spectrum both for $I=0$ and $I=2$. 
For the uncorrelated case, the spectrum has a peak at $E=1.54$ MeV, that 
is twice the single-particle resonance energy, 0.77 MeV. 
With the pairing interaction between the valence neutrons, the peak 
energy is shifted towards lower energies. The energy shift is larger in $I=0$ 
than in $I=2$. That is, the peak in the spectrum appears at 
$E=0.148$ MeV ($\Delta E= -1.392$ MeV) for 
$I=0$ and at $E=1.354$ MeV ($\Delta E=-0.186$ MeV) 
for $I=2$. 

As a matter of fact, it is quite natural that the 2$^+$ state appears at 
an energy slightly smaller than the unperturbed energy if the three-body 
picture is reasonable. 
In  standard textbooks of nuclear 
physics (see {\it e.g.,} Refs.\cite{T93,C00}), 
it is shown that the energy shift due to a pairing residual interaction, 
$v_{\rm res}(\vec{r}_1,\vec{r}_2)=-g \,\delta (\vec{r}_1-\vec{r}_2)$, is 
evaluated for a single-$j$ orbit as, 
\begin{eqnarray}
\Delta E_I &=& \langle [jj]^{(IM)}|
-g \,\delta (\vec{r}_1-\vec{r}_2) |[jj]^{(IM)}\rangle, \\
&=&-gF_r\,\frac{(2j+1)^2}{8\pi}
\left(
\begin{array}{ccc}
j & j & I \\
1/2 & -1/2 & 0
\end{array}
\right)^2,
\end{eqnarray}
where $F_r$ is the radial integral of the 
four single-particle wave functions. 
If one applies this formula to the $^{26}$O nucleus and sets $j=d_{3/2}$, 
one obtains 
\begin{equation}
\Delta E_{I=0}=-\frac{16}{8\pi}gF_r\cdot \frac{1}{4},
\label{deltaE0}
\end{equation}
and
\begin{equation}
\Delta E_{I=2}=-\frac{16}{8\pi}gF_r\cdot \frac{1}{20}. 
\label{deltaE2}
\end{equation}
That is, $\Delta E_{I=0}/\Delta E_{I=2}$ =5, which is compared to the value 
of $\Delta E_{I=0}/\Delta E_{I=2}$ =7.48 obtained in the present 
three-body 
model calculation. Eqs. (\ref{deltaE0}) and (\ref{deltaE2}) predict 
$\Delta E_{I=2}=-0.278$ MeV for the value $\Delta E_{I=0}=-1.392$ 
MeV of the three-body model prediction. 
Even though  
$\Delta E_{I=0}/\Delta E_{I=2}$ in the three-body model somewhat deviates from the simple estimates 
of Eqs. (\ref{deltaE0}) and (\ref{deltaE2}) 
due to the many-body continuum  effects, 
the small  energy shift for the 2$^+$ state 
can be well understood by these formulas derived for 
the single-$j$ model with the residual pairing interaction. 
The single-$j$ model indicates that the 2$^+$ state in $^{26}$O 
shows a typical spectrum 
governed by a pairing residual interaction and that 
the 2$^+$ energy never exceeds the unperturbed 
energy.  Thus it must be smaller than 1.54 MeV 
for the $^{26}$O nucleus since the single-particle resonance energy of $d_{3/2}$ state 
is 0.77 MeV in $^{25}$O. 

The lower panel of Fig. 1 shows the decay energy spectrum obtained by 
taking the linear superposition of the $I=0$ and $I=2$ contributions, 
that is, 
\begin{equation}
\frac{dP}{dE}=(1-\alpha)\frac{dP_{I=0}}{dE}+\alpha\frac{dP_{I=2}}{dE}. 
\end{equation}
The actual value of $\alpha$ would 
depend on the details of the wave function of $^{27}$F as well as 
the reaction dynamics of the proton knock-out reaction of $^{27}$F 
with which the 
initial state of $^{26}$O was prepared in the experiment 
of Ref. \cite{LDK12}. 
Since it is beyond the scope of this paper to evaluate the value 
of $\alpha$ in the actual experimental conditions, here we 
arbitrarily take $\alpha$=0.1. 
For comparison, 
the figure also shows 
the pure $I=0$ component, which is the same as that shown in the upper panel. 
One can see that the experimental data are 
reproduced slightly better by 
mixing the $I=2$ component, although the error bars are large and one may not 
draw a definite conclusion. 

The ground state energy 
may be slightly overestimated 
in this calculation 
in relation to the recent life time measurement of 
the $^{26}$O nucleus \cite{GMZ13}. We therefore repeat the same calculation 
by adjusting the $v_\rho$ parameter in the pairing interaction, 
Eq. (\ref{vnn}), so that the ground state energy becomes 5 keV, 
instead of 148 keV. 
If one uses this interaction, the 2$^+$ state in $^{26}$O appears at 
$E=1.338$ MeV. This value is similar to the previous result, 
$E=1.354$ MeV, and we thus conclude that the energy of the 2$^+$ 
state is much less sensitive to the $nn$ interaction as compared 
to the 0$^+$ state. 
This implies that the 2$^+$ state of $^{26}$O should definitely 
appear around $E=1.3$ MeV as long as the three-body picture is correct. 

In summary, we have discussed the $2^+$ state of $^{26}$O using the 
three-body model of $^{24}$O+$n$+$n$. We have shown that the 2$^+$ state 
appears at around $E=1.35$ MeV and this value does not change much 
even if we change the $nn$ interaction to vary the ground state energy 
from 150 keV to 5 keV. 
This 2$^+$ energy is close to, but slightly smaller than, 
the unperturbed energy, 
$E=1.54$ MeV, and 
thus the energy shift from the unperturbed energy is much smaller 
than the energy shift for the 0$^+$ state. 
We have argued that this 
is a typical spectrum 
well understood by the single-$j$ model with the pairing residual interaction. 
Many shell model calculations have predicted the excitation energy of the 2$^+$ state in $^{26}$O 
in the opposite trend, that is, they have predicted a higher energy 
than the unperturbed energy. 
It is desperately desired to confirm 
the energy of 2$^+$ state experimentally in order to clarify the validity of nuclear models 
and effective interactions in nuclei on and beyond the neutron drip-line.

\bigskip

We thank Y. Kondo and T. Nakamura for useful discussions. 
We also thank T. Suzuki for discussions on shell model calculations for the 
$^{26}$O nucleus. 
This work was supported by 
JSPS KAKENHI Grant Numbers 
25105503 and 26400263.

\end{document}